\documentstyle[prb,aps,epsf,twocolumn]{revtex}
\begin{document}
\draft

\title{Tight-binding molecular dynamic study of silver clusters}

\author{Jijun Zhao $^{\ast}$}

\address{Department of Physics and Astronomy, University of North Carolina at
Chapel Hill, Chapel Hill, NC 27599, USA.\\
International Centre for Theoretical Physics, P.O.Box 586, Trieste 34100,
Italy \\
}
\date{\today}
\maketitle
\begin{abstract}

Tight-binding molecular dynamics (TBMD) is used to study the structural and
electronic properties of silver clusters. The ground state structures of Ag
clusters up to 21 atoms are optimized via TBMD combined with
genetic algorithm (GA). The detailed comparison with {\em ab initio} results 
on small Ag$_n$ clusters ($n=3-9$) proves the validity of the tight-bind model. 
The clusters are found to undergo a transition from  ``electronic order'' to 
``atomic order'' at $n=10$. This is due to $s$-$d$ mixing at such size. The size
dependence of electronic properties such as density of states (DOS), $s$-$d$
band separation, HOMO-LUMO gap, and ionization potentials are discussed. 
Magic number behavior at Ag$_2$, Ag$_8$, Ag$_{14}$, Ag$_{18}$, Ag$_{20}$ is
obtained, in agreement with the prediction of electronic ellipsoid shell
model. It is suggested that both the electronic and geometrical shell exist
in the coinage metal clusters and they play a significant role in determining
cluster properties.

\end{abstract}

\section{Introduction}

The structural and electronic properties of metal clusters are currently a
field of intensive research both theoretically and experimentally \cite{1,2,3}.
The basic theoretical concept in the electronic structure of alkali metal
clusters is the shell model based on jellium sphere (or ellipsoid)
approximation \cite{1,2}, which has successfully interpreted the magic number
effect in Na$_n$ and K$_n$ clusters ($n=$2, 8, 20, 40, $\cdots$). As compared
to alkali-metal clusters, the application of electronic shell model to
coinage-metal clusters (Cu$_n$, Ag$_n$, Au$_n$) is somewhat questionable due
to the existence of inner $d$ electrons. Among the noble metal clusters,
Ag$_n$ is expected to exhibit the largest similarity to the alkali metal
clusters as the $4d$ orbitals in Ag atom are low-lying and act almost
like innershell core orbitals.

Experimental studies on silver clusters include mass-spectra \cite{4},
ionization potentials (IPs) \cite{5,6}, photoelectron spectra \cite{7,8,9},
electron spin resonance (ESR) \cite{10}, optical resonance absorption
\cite{11}, etc. In general, most of the cluster properties resemble the
predictions of shell model within one $s$ electron picture. But there are
still some experimental evidences for Ag$_n$ that are different from
alkali-metal clusters and cannot be understood via the shell model of $s$
electrons. For instance, the Mie resonance peak of silver clusters exhibit
blue shift with decreasing cluster radius \cite{11}, while red shift of Mie
resonance frequency is found for alkali-metal clusters. A direct comparison of
photoelectron spectra between Ag$_n$ and Na$_n$ display significant difference
that might be attributed to the $d$ orbitals \cite{8}. Therefore, it is
important to clarify the contribution of $4d$ electrons and $s-d$ interaction
in the silver clusters and the geometrical effect on the electronic properties
of the clusters.

Besides shell model, the metal clusters have been investigated by accurate
quantum chemical approaches based on molecular orbital theory \cite{3}. However,
such {\em ab initio} calculations on coinage metal clusters are quite time
consuming and limited in small size \cite{12,13,14,15}. Among those works,
the most detailed and comprehensive study of small neutral
silver clusters (Ag$_2$ to Ag$_9$) has been performed via configuration
interaction (CI) method with relativistic effective core potential (RECP)
\cite{14}. On the other hand, the electronic structures of several larger silver
clusters has been approximately described by a modified H\"uckel model
\cite{16}. However, all these studies are carried out for limited number of
structural candidates with symmetry constrain. An unbiased minimization for
the cluster ground state structure incorporated with electronic structure
calculations would be much more informative for understanding the interplay
between geometrical and electronic structure and testing the validity of
electronic shell model.

Up to now, the most reliable and accurate procedure in dynamic searching of
cluster equilibrium geometries is provided by Car-Parrinello (CP) method
with simulated annealing (SA) \cite{17}. But such kind of {\em ab initio} 
simulation is also limited in small size (about $n\leq 10$) in a
truly global optimization because of the rapidly increase in computational 
expense with cluster size. Among coinage metal clusters, the CP method has
been employed to study small Cu$_n$ clusters ($n=2-10)$ recently \cite{18},
but the corresponding investigation on silver and gold clusters is not
available so far. In recent years, tight-binding molecular dynamics has
been developed as an alternative to CP method in atomistic simulation for
larger scaled systems \cite{19}. As compared to {\em ab initio} methods, the
parameterized tight-binding (TB) Hamiltonian reduces the computational cost
dramatically. It is more reliable than classical simulation based on empirical
potential since the atomic motion is directly coupled with electronic structure
calculation at each steps. For transition-metal clusters, M.Menon and co-workers
have proposed a minimal parameter tight-binding scheme and used it to study
nickel and iron clusters \cite{20,21}. In this work, we shall introduce a
similar TB model for silver. By using the TBMD as a local minimization approach,
genetic algorithm (GA) \cite{22,23,24,25} is employed to search the global minimal
structure of Ag$_n$ clusters up to 21 atoms. The size dependence of relative 
stabilities, density of states (DOS), HOMO-LUMO  gaps, and ionization potentials 
of the clusters are calculated and compared with available experimental results.
The magic effect of electronic shell and the interplay between geometrical and 
electronic structure in silver clusters are discussed. 

\section{Theoretical method}

In the minimal parameter tight-binding scheme proposed by M.Menon {\em et al.}
\cite{20}, the total binding energy $E_b$ of transition-metal
atoms is written as a sum:
\begin{equation}
E_b=E_{el}+E_{rep}+E_{bond}.
\end{equation}
$E_{el}$ is the electronic band structure energy defined as the
sum of one-electron energies for the occupied states
\begin{equation}
E_{el}=\sum_k^{occ}\epsilon _k
\end{equation}
where energy eigenvalues $\epsilon _k$ can be obtained by solving orthogonal
$9n\times 9n$ TB Hamiltonian including $4d$, $5s$ and $5p$ electrons.
The repulsive energy $E_{rep}$ is described by a pair potential function $\chi
(r)$ of exponential form:
\begin{equation}
E_{rep}=\sum_i\sum_{j>i}\chi (r_{ij})=\sum_i\sum_{j>i}\chi _0e^{-4\alpha (r-d_0)}
\end{equation}
where $r_{ij}$ is the separation between atom $i$ and $j$, $d_0=2.89$\AA$~$ is the
bond length for the fcc bulk silver \cite{26}, $\alpha$ is taken to be
one-half of $1/d_0$ according to Ref.20.

In order to reproduce the cohesive energies of small clusters through bulk TB
hopping parameters, it is still necessary to introduce a bond-counting term
$E_{bond}$:
\begin{equation}
E_{bond}=-N[a(n_b/N)+b]
\end{equation}
Here the number of bonds $n_b$ are evaluated by summing over all bonds
according to cutoff distance $R_c$
\begin{equation}
n_b=\sum_i[{\rm exp}(\frac{d_i-R_c}\Delta )+1]^{-1}.
\end{equation}
It should be noted that only the first two terms $E_{el}$ and $E_{rep}$ in
Eq.(1) contribute to the interatomic forces in TBMD simulation, while the
$E_{bond}$ term is added after the relaxation has been achieved. However, for
metal clusters, this correction term is significant in distinguishing various
isomers at a given cluster size \cite{20}.

The $9n\times 9n$ TB Hamiltonian matrix is constructed with Slater-Koster
scheme, while the distance scaling of hopping integrals
$V_{\lambda \lambda ^{\prime }\mu }$ is taken as the Harrison's
universal form \cite{27}:
\begin{equation}
V_{\lambda \lambda ^{\prime }\mu }(d)=V_{\lambda \lambda ^{\prime }\mu
}(d_0) (\frac{d_0}{d})^{\tau +2}
\end{equation}
The parameter $\tau =0$ for $s$-$s$, $s$-$p$, $p$-$p$ interactions, $\tau =3/2$
for $s$-$d$ and $p$-$d$ interaction, $\tau =3$ for $d$-$d$ interaction.

In present, we have adopted the Slater-Koster hopping integrals
$V_{\lambda \lambda ^{\prime }\mu }(d_0)$ and the on-site orbital energy
from the values fitted to first principle APW band structure calculation of
bulk silver \cite{27}. Furthermore, to transfer the on-site orbital energies
levels from bulk calculation to atomic limit, a constant energy shift
$\Delta \epsilon $=-15.88 eV is applied on the on-site energies from Ref.27.
Such shift in on-site levels make the theoretical ionization potential of 
Ag$_n$ clusters quantitatively comparable to experimental values. The repulsive 
potential parameter $\chi _0$ is fitted for experimental bond length 2.48 \AA$~$ 
of silver dimer \cite{28}. The bond-counting terms $a$, $b$ are chosen to
reproduce the {\em ab initio} binding energy for small clusters
Ag$_3$, Ag$_4$, Ag$_5$ \cite{12,14,15}. All the parameters used in our
calculation are listed in Table I. These empirical parameters can describe
both bulk phase and dimer of silver with an acceptable accuracy.
The cohesive energy 2.75 eV, equilibrium interatomic distance 2.88 \AA$~$
of fcc solid silver from TB model are close to the corresponding
experimental value 2.95 eV and 2.89 \AA $~$ respectively \cite{26}.
The vibrational frequency and binding energy calculated for silver dimer at 
equilibrium distance is 125.5 cm$^{-1}$ and 1.25 eV, in reasonable agreement with
experimental data of 192.4 cm$^{-1}$ and 1.66 eV \cite{29}.

\begin{table}
Table I. Parameters of TB model for silver used in this work.\\
\begin{center}
\begin{tabular}{cccc}
$\epsilon_s$&$\epsilon_p$&$\epsilon_{d, xy}$&$\epsilon_{d, x^2-y^2}$ \\
   -6.452 eV & -0.447 eV  &  -14.213 eV     &  -14.247 eV \\
\end{tabular}
\begin{tabular}{ccccc}
$V_{ss\sigma}$&$V_{sp\sigma}$&$V_{pp\sigma}$&$V_{pp\pi}$&$V_{sd\sigma}$ \\
  -0.895 eV   &   1.331 eV   &   2.14317 eV &  0.088 eV & -0.423 eV \\
\end{tabular}
\begin{tabular}{ccccc}
$V_{pd\sigma}$&$V_{pd\pi}$&$V_{dd\sigma}$&$V_{dd\pi}$&$V_{dd\delta}$ \\
-0.531 eV     & 0.207 eV  &  -0.429 eV   &  0.239 eV &  -0.046 eV \\
\end{tabular}
\begin{tabular}{ccccccc}
$d_0$ &$\alpha$ & $\chi_0$ & $a$ & $b$ & $R_c$ & $\Delta$ \\ 
2.89 \AA &  0.692\AA$^-1$ & 0.58 eV & -0.16 eV & 0.59 eV & 3.5 \AA & 0.1 \AA
\\
\end{tabular}
\end{center}
\end{table}

The determination of lowest energy structures of clusters is performed by 
a genetic algorithm (GA) \cite{22,23,24,25}. The essential
idea of this novel global optimization strategy is to mimic the Darwinian 
biological evolution process in which only the fittest candidate can survive. 
Some pioneering works have demonstrated the impressive efficiency of GA in
searching the global minima of clusters as compared to standard simulated 
annealing. At beginning, we generate a number $N_p$ of initial configurations 
by random ($N_p=4-16$, depending upon cluster size). Any two candidates in 
this population can be chosen as parents to generate a child
cluster through mating process. In the end of mating procedure, mutation
operation is allowed to apply on the geometrical structure of child cluster
with 30$\%$ possibility. The child cluster from each generation is
relaxed by TBMD quenching of 300-500 MD steps with $\Delta t=5fs$. Then the
locally minimized child is selected to replace its parent in the population
if it has different geometry but lower binding energy. Typically, 100$-$150 GA
iterations is sufficient to ensure a truly global search up to $n=21$. The
detailed description about the practical operation of GA is given
elsewhere \cite{25,30}.

\section{Structures and stabilities of sliver clusters}

\subsection{Structures of small Ag$_n$ with $n\leq 9$}

By using the combined GA-TBMD strategy, we have explored the global minimal
structures of Ag$_n$ clusters up to 21 atoms. These ground state structures are 
shown in Fig.1 ($4\leq n\leq 9$) and Fig.2 ($10\leq n\leq 21$).
In Table II, the structural parameters (bond lengths and bond angles), binding
energies and ionization potentials of the small Ag$_n$ clusters ($n=3-9$) in 
ground state and some metastable structures are compared with
accurate quantum chemistry calculations \cite{12,13,14,15}. The lowest
energy structures found for most of the clusters coincide well with intensive CI
calculations in Ref.14 and the other works \cite{12,13,15}. The calculated
cluster properties of ground state and metastable isomers agree with {\em ab
initio} results quantitatively. As shown in Table I, the TB bond lengths are
typically within 0.05$-$0.15 \AA$~$ according to the {\em ab initio}
values. The average deviation of binding energy per atom and ionization
potentials from this work to {\em ab initio} calculations \cite{14} is
about 0.13 eV and 0.30 eV respectively.

\begin{figure}
\centerline{
\epsfxsize=3.0in \epsfbox{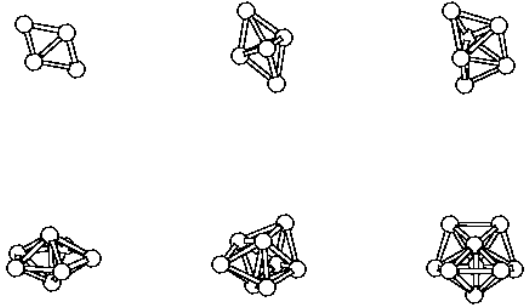}
}
\caption{Lowest-energy structures for Ag$_n$ ($n=4-9$) clusters.
}
\end{figure}

For silver trimer, isosceles triangle is about 0.05 eV lower in energy than
equilateral triangle and 0.84 eV lower than the linear isomer. ESR experiments
on Ag$_3$\cite{9} supports the isosceles triangle structure with C$_{\rm 2v}$
symmetry. In the case of Ag$_4$, planar rhombus is lower in energy than a
relaxed tetrahedron by $\Delta E$=0.31 eV although the tetrahedron has higher
coordination and has been predicted as most stable structure by using a
classical many-body potential \cite{30}. This discrepancy demonstrates the 
importance of incorporating the electronic structure in determining cluster 
geometries. The lowest energy structure found for Ag$_5$ is a compressed 
trigonal bipyramid, which has lower energy ($\Delta E$=0.17 eV)
than a planar capped rhombus. In previous studies, the planar structure has 
been obtained as ground state \cite{12,14,15} but the energy difference 
between these two isomers is rather small ($\Delta E$=0.31 eV in Ref.12 and 
$\Delta E$= 0.003 eV in Ref.14). It is noted that the experimental ESR 
spectra of Ag$_5$ can be interpreted by a geometrical structure of deformed 
trigonal bipyramid. Two isoenergitic structures, a bicapped tetrahedron and
a pentagonal pyramid are found for Ag$_6$, with $\Delta E$=0.05 eV. The
bicapped tetrahedron is more stable but this conclusion depends sensitively
on the choice of empirical parameters. In Ref.14, these two structures are also
found to be very close in energy ($\Delta E=$ 0.06 eV) but the pentagonal
pyramid is ground state. According to the theoretical HOMO-LUMO gap of these
two isomers, we suggest that the bicapped tetrahedron is a better candidate
since its HOMO-LUMO gap (1.05 eV) is much smaller than that obtained for
pentagonal pyramid (2.67 eV), whereas experimental gap is about 0.34 eV
\cite{7}. The pentagonal bipyramid is obtained as lowest energy structure
for Ag$_7$. The tricapped tetrahedron is a locally stable isomer with
$\Delta E=$0.48 eV, while the $\Delta E$ in Ref.14 for the same isomer is
0.41 eV. For silver octamer, a bicapped octahedron is our ground state
structure, which is also found for Cu$_8$ \cite{18}. This near-spherical
configuration can be understood by electronic shell model. The closure of
electronic shell at $n=8$ might give rise to a spherical charge density
distribution, which favors the formation of spherical atomic arrangement.
The tetracapped tetrahedron is predicted as metastable isomer in Ref.14 but
it is unstable upon relaxation in our simulation. Another spherical-like
structure, square antiprism (D$_{\rm 4d}$) is found as a local stable isomer
with $\Delta E$=0.99 eV. For Ag$_9$, the ground state structure is a
bicapped pentagonal bipyramid. Its energy is lower than that of the tricapped
trigonal prism (C$_{\rm 3v}$) by 0.59 eV and than that of capped square
antiprism (C$_{\rm 2v}$) by 1.01 eV. In Ref.14, bicapped pentagonal bipyramid
is also found as ground state and the energy difference $\Delta E$ for the two
structural isomers is 0.73 eV and 0.22 eV respectively.

\begin{table}
Table II Comparison of structural properties (bond length, bond angle),
average binding energies $E_b/n$, and vertical ionization potentials (IP) of
small Ag$_n$ ($n=3-9$) clusters with {\em ab initio} calculations
\cite{12,13,14,15}. The definition of structural parameters $r$, $\alpha$,
$h$, etc. for smaller Ag$_{3-5}$ clusters is chosen according to Ref.[14]; the
bonds for Ag$_{6-9}$ are defined by their lengths in Ref.[14] in a declining
sequence. $a$ denotes our present tight-binding calculation. $b$ to $e$ are
previous {\em ab initio} calculations based on relativistic effective core
potential configuration (RECP): $b$-modified coupled pair function (MCPF)
\cite{12}; $c$-multireference singles plus doubles configuration (MRSDCI)
\cite{13}; $d$-configuration interaction (CI) \cite{14}; $e$-relativistic
effective core potential density functional theory (RECP-DFT) \cite{15}.

\begin{center}
\begin{tabular}{ccccc}
\multicolumn{5}{c}{Ag$_3$, obtuse triangle (${\rm C_{2v}}$)} \\
    & $r$ (\AA ) & $\alpha$ (deg) & $E_b/n$ (eV)& IP (eV) \\ \hline
$a$ & 2.659      &   66.8         & 0.82        & 5.65    \\
$b$ & 2.709      &   69.2         & 0.80        & 5.59    \\
$c$ & 2.720      &   63.7         & 0.90        & 5.90    \\
$d$ & 2.678      &   69.1         & 0.86        & 5.74    \\
$e$ & 2.627      &   70.4         & 0.84        &  --     \\
\end{tabular}
\end{center}

\begin{center}
\begin{tabular}{ccccc}
\multicolumn{5}{c}{Ag$_4$, rhombus(${\rm D_{2h}}$)} \\
    & $r$ (\AA ) & $\alpha$ (deg) & $E_b/n$ (eV)& IP (eV) \\ \hline
$a$ & 2.731      &   56.6         & 1.21        & 6.86 \\
$b$ & 2.862      &   57.6         & 1.11        & 6.54 \\
$c$ & 2.870      &   55.5         & 1.83        & 6.40 \\
$d$ & 2.800      &   56.4         & 1.20        & 6.60 \\
$e$ & 2.740      &   57.2         & 1.11        & ---  \\
\end{tabular}
\end{center}

\begin{center}
\begin{tabular}{cccccc}
\multicolumn{6}{c}{Ag$_5$, deformed trigonal bipyramid (${\rm C_{2v}}$)}\\
    &$r$ (\AA)& $\alpha$ (deg) & $h/2$ (\AA) &$E_b/n$(eV)&IP (eV)\\ \hline
$a$ & 2.749     &  67.5        &  2.34       & 1.38      & 5.88  \\
$b$ & 2.858     &  65.8        &  2.39       & 1.16      & ---   \\
$d$ & 2.709     &  67.8        &  2.33       & 1.28      & 5.95  \\
\end{tabular}
\end{center}

\begin{center}
\begin{tabular}{ccccccc}
\multicolumn{7}{c}{Ag$_5$, planar capped rhombus (${\rm C_{2v}}$)}\\
    &$r_1$ (\AA)&$r_2$ (\AA)&$r_3$ (\AA)&$r_4$ (\AA)&$E_b/n$(eV)&IP (eV) \\ \hline
$a$ & 2.851     &  2.736    & 2.740     & 2.668     & 1.32   & 6.20  \\
$b$ & 2.842     &  2.842    & 2.842     & 2.842     & 1.22   & 6.18  \\
$d$ & 2.812     &  2.801    & 2.760     & 2.759     & 1.28   & 6.20  \\
\end{tabular}
\end{center}

\begin{center}
\begin{tabular}{cccccccc}
\multicolumn{8}{c}{Ag$_6$, bicapped tetrahedron (${\rm C_{2v}}$)}\\
    &$r_1$ (\AA)&$r_2$ (\AA)&$r_3$ (\AA)&$r_4$ (\AA)&$r_5$ (\AA) &$E_b/n$(eV)& IP(eV) \\ \hline
$a$ & 2.931   &  2.875  & 2.766   & 2.653  & 2.661 & 1.65  & 6.71\\
$d$ & 2.976   &  2.859  & 2.783   & 2.751  & 2.672 & 1.49  & 6.23\\
\end{tabular}
\end{center}

\begin{center}
\begin{tabular}{ccccc}
\multicolumn{5}{c}{Ag$_6$, pentagonal pyramid (${\rm C_{5v}}$)}\\
    &$r_1$ (\AA)& $r_2$ (\AA)& $E_b/n$(eV) & IP (eV) \\ \hline
$a$ & 2.984     &  2.539       &  1.65     &  7.92   \\
$d$ & 2.828     &  2.740       &  1.50     &  7.00   \\
\end{tabular}
\end{center}

\begin{center}
\begin{tabular}{ccccc}
\multicolumn{5}{c}{Ag$_7$, pentagonal bipyramid (${\rm D_{5h}}$)} \\
    &$r_1$ (\AA)& $r_2$ (\AA)& $E_b/n$(eV) & IP (eV) \\ \hline
$a$ & 2.879     &  2.858       &  1.87     & 5.95    \\
$d$ & 2.815     &  2.806       &  1.71     & 5.91    \\
\end{tabular}
\end{center}

\begin{center}
\begin{tabular}{ccccccc}
\multicolumn{7}{c}{Ag$_8$, bicapped octahedron (${\rm D_{2d}}$)}\\
    &$r_1$ (\AA)&$r_2$ (\AA)&$r_3$ (\AA)&$r_4$ (\AA)&$E_b/n$(eV)& IP(eV) \\ \hline
$a$ & 3.140     &  2.812    & 2.941     & 2.661     & 2.03  & 7.10\\
$d$ & 2.973     &  2.812    & 2.804     & 2.735     & 1.80  & 6.80\\
\end{tabular}
\end{center}

\begin{center}
\begin{tabular}{cccccc}
\multicolumn{6}{c}{Ag$_9$, bicapped pentagonal bipyramid  (${\rm C_{2v}}$)}\\
    &$r_1$ (\AA)&$r_2$ (\AA)&$r_3$ (\AA)&$r_4$ (\AA) $r_5$ (\AA) &\\ \hline
$a$ &2.989& 2.936& 2.993 & 2.888 & 2.856 \\
$d$ &2.934& 2.887& 2.881 & 2.836 & 2.786 \\ \\ \hline
    & $r_6$ (\AA )&$r_7$ (\AA )&$E_b/n$(eV)& IP(eV) & \\
$a$ & 2.849 & 2.745 & 2.09 & 5.74 & \\
$d$ & 2.766 & 2.752 & 1.77 & 5.10 & \\
\end{tabular}
\end{center}
\end{table}

From the above discussions for small silver clusters, we find
that the agreement between TB model and {\em ab initio} calculations
is satisfactory, particularly considering the simplicity in the present
tight-binding scheme. Therefore, in the next part, we shall use this model
to study larger clusters with $n\geq 10$ for which the global minimization
with {\em ab initio} molecular dynamics is much more expensive.

\subsection{Structures of Ag$_n$ with $10\leq n\leq 21$ }

\begin{figure}
\centerline{
\epsfxsize=3.0in \epsfbox{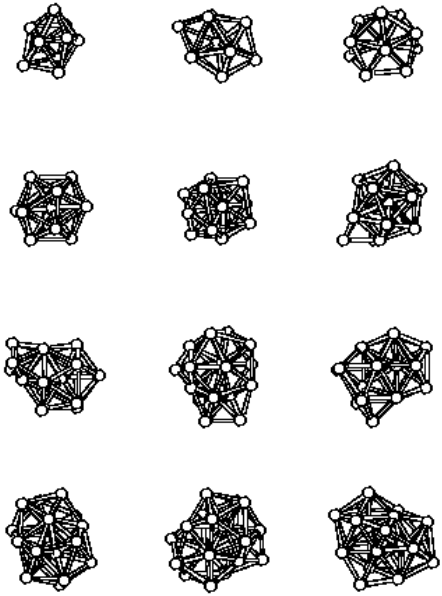}
}
\caption{Lowest-energy structures for Ag$_n$ ($n=10-21$) clusters. See text
for description of structures.
}
\end{figure}

The lowest energy structure of Ag$_n$ ($n=10-21$) obtained from GA-TBMD
simulation is shown in Fig 2. The most stable structure of Ag$_{10}$ is
a deformed bicapped square antiprism (D$_{\rm 4d}$), which is similar to that
found for Cu$_{10}$ \cite{18}. Starting from Ag$_{11}$, the ground
state structures of Ag$_n$ clusters are based on icosahedral packing,
except for Ag$_{14}$. Many other capped polyhedral structures are obtained as
local isomers for Ag$_n$ with $n=$10-21 but it is not necessary to describe
them herein. As shown in Fig.2, the structures of Ag$_{11}$, Ag$_{12}$ are
the uncompleted icosahedron with lack of one or two atoms. An Jahn-Teller
distorted icosahedron is then formed at Ag$_{13}$.
Following the icosahedral growth sequence, the lowest energy structures of
Ag$_{15}$, Ag$_{16}$, Ag$_{17}$ is the icosahedron capped with 2, 3, 4 atoms
respectively. The capped single icosahedron transits into an truncated double
icosahedron at Ag$_{18}$ and a complete double icosahedron at the Ag$_{19}$.
Base on the double icosahedron structure, the structures of Ag$_{20}$ and
Ag$_{21}$ is formed by one and two atoms capped on that of Ag$_{19}$. However,
an exception occur at Ag$_{14}$, for which we found a fcc-like packing with
4-1-4-1-4 layered structure. This ellipsoid structure is more stable than a
spherical capped icosahedron structure by 0.03 eV. It is
worth noted that 14 is a magic size predicted by a ellipsoid shell model
\cite{1,30}.

\subsection{Crossover from ``electronic order'' to ``atomic order''}

The concept of ``particle order'' (or ``atomic order'') and ``wave order'' (or
``electronic order'') in cluster physics \cite{32} have been introduced
in order to explain the magic number effect in the inert gas clusters
with atomic shell and alkali metal clusters with electronic shell.
The noble metal cluster is a mixture of the atomic core involving the
relatively localized $d$ electrons and the more delocalized $s$ valence
electrons. Therefore, it might be a intermediated system from these two
extreme limit and exhibit features come from both the two orders. The
equilibrium structures of small silver clusters (Ag$_3$$-$Ag$_{9}$) from
our calculation are similar to those of alkali-metal clusters \cite{3}.
In contrast, the icosahedron growth sequence is obtained for clusters
starting from Ag$_{11}$ that has also been found for noble gas clusters
\cite{33}. In smaller silver clusters, the $5s$ valence electrons are dominant
in determining the cluster property and the $d$ states are significantly
lower-lying and contribute much less to the cluster bonding. Therefore, these
small clusters should exhibit certain alkali-metal-like behavior in both
structural and electronic aspects. As the clusters size increase, the
contribution of $d$ electrons to cluster bonding become more and more
important. The bonding energy from $d$
electrons is roughly related to the $d$ band width \cite{34}, which is
approximately proportional to the square root of the average coordination
number of the clusters \cite{35}. Consequently, the clusters tend to adopt the
more compact structures such as icosahedron, which is similar to noble gas
clusters \cite{33}. The switch of structural pattern from alkali-metal like to
noble gas like at around n=10 can be identified as a crossover from
``electronic order'' towards ``atomic order'' in the silver
clusters. This alternation is related to the overlap of $4d$ and $5s$
electronic states which we will discussed latter. However, our further study
show that the shell structure of $s$ electrons still dominates the electronic
properties such as IPs, HOMO-LUMO gaps of the silver clusters although the
geometrical structures has taken ``atomic order''. We argue that the
``electronic order'' and ``atomic order'' can coexist in coinage metal clusters.

\subsection{Size dependence of relative stabilities}

The second differences of cluster binding energies defined by
$\Delta_2 E (n)=E_b(n+1)+E_b(n-1)-2E_b(n)$ is calculated and plotted in Fig.3.
In cluster physics, it is well known that the $\Delta_2 E (n)$ is a
sensitive quantity which reflects the stability of clusters and can be
directly compared to the experimental relative abundance. Three major
characteristics can be found in the Fig.3: (i) even-odd alternation of
$\Delta_2 E (n)$ with $n=2-6, 15-21$; (ii) particular high peak at Ag$_8$,
Ag$_{18}$; (iii) other maxima at odd size like Ag$_{13}$ and Ag$_{11}$. The
first effect can be related to the even-odd oscillation of HOMO energy and
HOMO-LUMO gap in silver clusters, which is due to electron pairing effect.
The articular stable clusters such as Ag$_8$, Ag$_{18}$ corresponds to the
magic number in electronic shell model. However, the even-odd oscillation in
$\Delta_2 E (n)$ from Ag$_{10}$ to Ag$_{14}$ and the maximum at magic size
Ag$_{20}$ have not been observed in our calculation. In stead, some odd-sized
cluster as Ag$_{11}$, Ag$_{13}$ become maxima in Fig.3. These phenomena can be
attributed to the geometrical effect. The closing of geometrical shell of
icosahedron at Ag$_{13}$ will enhance the stability of such clusters and
reduce the relative stability of their neighboring clusters.

\begin{figure}
\centerline{
\epsfxsize=3.0in \epsfbox{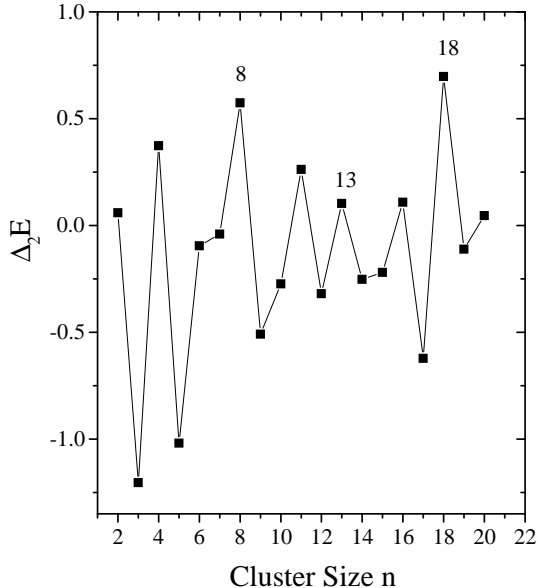}
}
\caption{Second differences of cluster binding energies $\Delta E(n)=
[E_{b}({\rm Ag}_{n-1})+E_{b}({\rm Ag}_{n+1})]-2E_{b}({\rm Ag}_n)$ as a
function of cluster size $n$ for $n=2-21$. Both electronic shell effect at
$n=2$, 8, 18 and geometrical shell effect at Ag$_{13}$ can be identified.
See text for details.}
\end{figure}

The simultaneous appearance of those three features in the $\Delta_2 E (n)$
demonstrates that the structure and stability of a silver cluster is
determined by both electronic structure and atomic configuration. Either
electronic or geometrical effect is enhanced if the corresponding shell
structure is completed. This argument is supported by a experimental
probe of geometrical and electronic structure of copper clusters \cite{36}.
They found both jellium-like electronic behavior and icosahedral geometrical
structure in copper clusters. In a experimental studies of mass spectra of
ionized silver clusters \cite{4}, dramatic even-odd oscillation as well as
substantial discontinuities at electronic magic number 8, 20 ($n=9$, 21 for
cationic clusters) are found. The discrepancy between present theoretical
result and experiment may be partially attributed to the effect of ionization
on the cluster stability. Since the experimental mass spectra distribution is
recorded for ionized clusters Ag$_n^+$, it is possible that the charging on
the cluster can significantly  alter the geometrical and electronic structure
of the cluster \cite{3,14}.

\section{Electronic properties vs cluster size}

\subsection{Size evolution of electronic band}

We investigated the cluster electronic properties via the electronic density of
states (DOS). In Fig.4, we present the total $spd$ electronic DOS for Ag$_2$,
Ag$_8$, Ag$_{13}$ along with bulk DOS of fcc crystal
silver from TB calculation in reciprocal space. Generally, the total DOS is
composed by the relatively compact $d$ states and the more expanded $sp$
states. In smallest
clusters such as Ag$_2$, the $d$ and $sp$ bands are clearly separated. The
$sp$ states shows discrete peaks originated from symmetrical splitting of
atomic orbital levels, while the $d$ band is low-lying and considerably
narrower than the bulk $d$ band. In contrast to even-odd behavior and shell
occupation of $s$ electrons, the evolution of $d$ states from smallest
clusters towards bulk solid is a monotonic broaden of band width.
As the cluster size increases, both $d$ and $sp$ levels gradually broaden,
shift, overlap with each other, and finally come into being bulk electronic
band. The DOS of Ag$_8$ still has molecular-like some discrete $sp$ peaks
but these electronic spectra peaks tend to overlap and form continuous band.
In Ag$_{13}$, the $sp$ states have developed into several subbands and the
$d$ band has overlapped with $sp$ states. Although the
DOS of Ag$_{13}$ is roughly similar to the bulk limit, the width of
$d$ band is still considerably narrower than the bulk $d$ band width and the
fine structure of $sp$ electronic spectra is somewhat different from bulk
$sp$ band. This fact suggests that the bulk-like behavior emerge at around
Ag$_{13}$. We have also studied the electronic states of Ag$_{55}$ with
icosahedral and cuboctahedral structures by using present tight-binding
scheme with local minimization. The DOS for both of them are much closer to
bulk band. In a previous experimental study of photoelectron spectra of silver
clusters up to 60 atoms \cite{8}, it is found that the ultraviolet
photoelectron spectroscopy (UPS) of smallest Ag$_n$, i.e., $2\leq n\leq 10$ is
different from bulk UPS and changes sensitively on cluster size. The size
dependent variation of UPS for Ag$_n$ with $n<10$ becomes more gradual and
the UPS of Ag$_{60}$ is already very close to that of solid silver.

\begin{figure}
\centerline{
\epsfxsize=3.0in \epsfbox{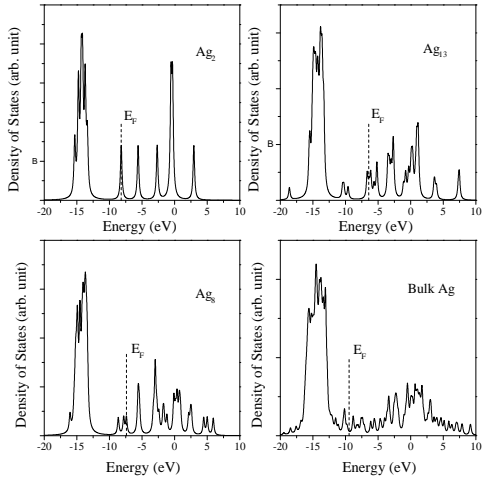}
}
\caption{Density of states (DOS) of Ag$_n$ ($n=2$, 8, 13) clusters vs
cluster as well as bulk DOS in fcc crystalline (with Gaussian broaden of
0.02 eV).}
\end{figure}

To further clarify the size evolution of the overlap of $d$ and $sp$ bands in
the small silver clusters, we have examined the energy separation $\Delta_{sd}$
between the highest molecular orbitals belong to $d$ states and lowest
molecular orbitals from $s$ states. The calculated $\Delta_{sd}$ decrease
rapidly from 5.20 eV for Ag$_2$, to 1.69 eV for Ag$_5$ and then to 0.10 eV for
Ag$_{8}$. Finally, the $d$ and $sp$ band merge in Ag$_n$ clusters with
$n\geq 9$. The overlap between $s$ and $d$ band is believed to induce the
icosahedral growth sequence starting from Ag$_{11}$ and weaken the even-odd
oscillation in HOMO-LUMO gaps and IPs of Ag$_n$ clusters with $n>10$. However,
the overlap between $sp$ and $d$ states is small and the cluster HOMO is still
located in the $s$-like electronic states. Consequently, the contribution from
$d$ electrons in Ag$_n$ with $n>9$ shows more importance to cluster bonding
although the electronic behavior close to Fermi energy such as HOMO-LUMO gap,
ionization potentials is still dominated by $s$ orbitals.

\subsection{HOMO-LUMO gap}

\begin{figure}
\centerline{
\epsfxsize=3.0in \epsfbox{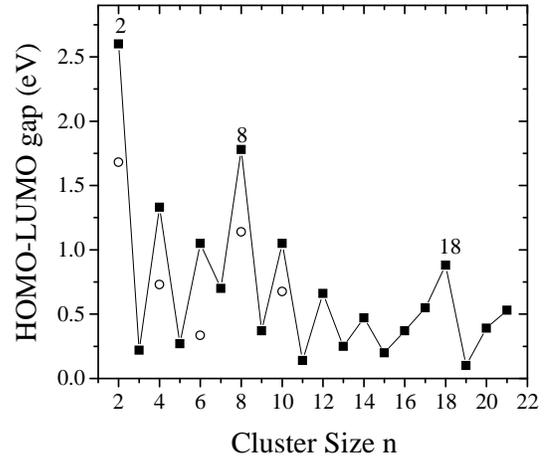}
}
\caption{HOMO-LUMO gap $\Delta$ (eV) vs cluster size $n$. The
theoretical values are labeled by solid square connected with solid line
and the experimental values in Ref.[7] are labeled by open circles.
Electronic shell effect for $n=2$, 8, 18 can be clearly identified.
}
\end{figure}

An important electronic property of a cluster is the gap between highest
occupied molecular orbital (HOMO) and lowest unoccupied molecular orbital
(LUMO). In the case of magic cluster, the closure of electronic shell shall
manifest itself in particularly large HOMO-LUMO gap. This effect was
demonstrated experimentally for small even-sized silver and copper
clusters \cite{7} and theoretically for copper clusters \cite{37,38}.
The theoretical HOMO-LUMO gap of Ag$_n$ ($n=2-21$) along with experimental gap
of small clusters Ag$_n$ ($n=2$, 4, 6, 8, 10) \cite{7} are shown in
Fig.5. Even-odd oscillation up to Ag$_{16}$ as well as the particularly large
HOMO-LUMO gap at Ag$_2$, Ag$_8$, and Ag$_{18}$ are obtained. As compared to
the experimental results for small Ag$_n$ with even size, the present TB
calculation has systematically overestimated the HOMO-LUMO electronic gap by
about 0.5 eV. But the size dependent variation of experimental gaps and
magic effect in HOMO-LUMO gaps at $n=$2, 8 are well reproduced.
The even-odd alternation for $n\geq 16$ and magic effect of Ag$_{20}$
have not been obtained in our calculation. We suggest these are probably
due to the geometrical effect, since the HOMO-LUMO gap of cluster depends
sensitively on cluster structure \cite{37}.
In a previous study of HOMO-LUMO gaps of copper clusters \cite{37}, the maxima
gap at Ag$_8$ and Ag$_{20}$ is found but the even-odd alternation of electronic
gap and magic effect for Ag$_{18}$ have not been obtained.

\subsection{Ionization potential}

The vertical ionization potentials (IPs) of clusters are evaluated from the
highest occupied molecular orbital (HOMO) energy of neutral clusters according
to Koopman's theorem. In Fig.6, the calculated IPs of Ag$_n$ up to $n=21$ is
compared with the IP values measured by C.Jackschath \cite{5}, the prediction
by metallic spherical droplet model \cite{39}, and the size dependent HOMO
level (in arbitrary units) of alkali-like metal clusters by Clemenger-Nilsson
ellipsoid shell model \cite{1,31}. In comparison with experimental values in
Ref.[5], the present TB calculation has almost reproduced the size dependence
of IPs for silver clusters up to 21 atoms except that theoretical calculation
has overestimated the IP values of some magic clusters such as Ag$_2$, Ag$_8$,
and Ag$_{18}$. Two important size dependent behaviors are found in Fig.6:
(i) dramatic even-odd alternations where clusters with
even number of $s$ valence electrons have higher IPs than their immediate
neighbors; (ii) particular higher IP values at the magic clusters such as
Ag$_2$, Ag$_8$, Ag$_{18}$, Ag$_{20}$ and probably Ag$_{14}$.
The even-odd variations can be attributed to electron pairing effect.
Odd(even)-sized clusters have an odd(even) total number of $s$ valence
electrons and the HOMO is singly(doubly) occupied. The electron in a doubly
occupied HOMO feels a stronger effective core potential since the electron
screening is weaker for the electrons in the same orbital than for inner shell
electrons. Therefore, the binding energy of a valence electron in a cluster of
even size cluster is larger than that of odd one. It is also
interesting to note that the size dependence of IPs from TB model is
almost in full accordance to Clemenger-Nilsson shell model. The magic effect at
Ag$_2$, Ag$_8$, Ag$_{14}$, Ag$_{18}$, and Ag$_{20}$ predicted by electronic
shell model is well reproduced even though the cluster geometries and $s$-$d$
interaction has been considered. On the other hand, one can found that
both the theoretical and experimental IP values for silver clusters can be
roughly described by a classical electrostatic model which take the cluster
as a metallic spherical droplet. This fact further suggests that the electronic
behavior of silver clusters are predominantly $s$-like and the cluster can be
approximated to be a structureless jellium sphere with shell-like
electronic levels.

\begin{figure}
\centerline{
\epsfxsize=3.0in \epsfbox{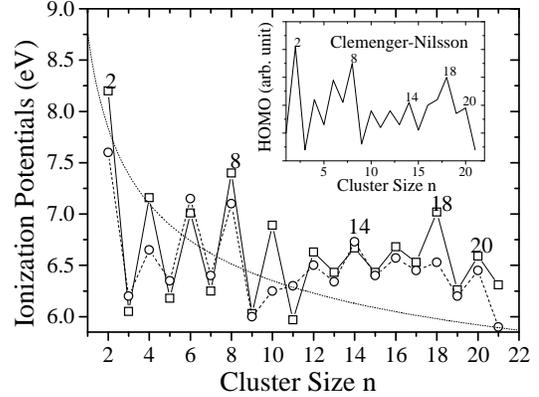}
}
\caption{Vertical ionization potentials (IPs) vs cluster size $n$.
The measured IP values reported in Ref.[5] are labeled by open circles
connected with dashed line, the theoretical values from TB calculation are
labeled by open squares connected with solid line. Electronic shell closing
exhibit as pronounced drops of IP at $n=2$, 8, 14, 18, 20. See text for details.
}
\end{figure}

\section{Conclusion}

In conclusion, we have shown that the present TB model is able to describe
the geometrical and electronic structures of silver clusters. By using
a combined GA-TBMD technique, the lowest energy structures, binding energies,
electronic states, s-d separation, HOMO-LUMO gap, vertical ionization
potentials are obtained and compared with experiments. The main theoretical
results can be summarized in the following points:

(1) The structures of small silver clusters is determined by $s$ electrons
and similar to those of alkali-metal clusters. The contribution of $d$
electrons to cluster bonding become more important as cluster size exceed
10 atoms. The icosahedral growth pattern starts in the Ag$_n$ with $n \geq 11$,
which can be identified as ``atomic order'' clusters like noble gas clusters.

(2) The electronic and geometrical shell structure coexist in silver clusters
and take effect in the clusters simultaneously. The electronic shell effect on
cluster electronic properties has been found by present TB calculation, in
which the effect of geometrical structures and $d$ electrons
are directly incorporated. The silver clusters with closed electronic shell
($n=$2, 8, 14, 18, 20) show more pronounced electronic characteristics while
the geometrical effect is enhanced as the icosahedral shell completes at
Ag$_{13}$.

(3) Due to the pair occupation of $s$ valence electrons on molecular orbitals,
silver clusters show even-odd alternation in their relative stability,
HOMO-LUMO gap, ionization potential. However, the even-odd effects can be
disturbed by the $sd$ overlap or the geometrical structures of clusters.

(4) The density of electronic states of smaller silver cluster, e.g., $n<10$,
is composed by discrete $sp$ expanded band and a narrow $d$ band. The bulk-like
feature in DOS start at around Ag$_{13}$ and the bulk limit can be roughly
reached by $n=55$.

The present study shows that both the geometrical and electronic effects should
be considered in order to achieve a complete description of coinage clusters.
Therefore, {\em ab initio} molecular dynamics or TBMD are essential to
elucidate the interplay between geometrical and electronic structures of these
clusters. The further works should include the larger clusters and extend
the TB model to other transition metal elements.

\begin{acknowledgements}
This work is partially supported by the U.S. Army Research Office
(Grant DAAG55-98-1-0298). The author is deeply grateful to Dr. J.Kohanoff
and Dr. J.P.Lu for stimulating discussion and critical reading of manuscript.
\end{acknowledgements}

\vspace{0.5cm}
\noindent
$^{\ast}$ E-mail: zhaoj@physics.unc.edu

\end{document}